\ifx\onecol\undefined
\documentclass[journal,twocolumn]{IEEEtran}
\else
\documentclass[12pt,draftclsnofoot,journal,onecolumn]{IEEEtran}
\fi

\usepackage{amsmath,amssymb,amsfonts}
\usepackage{bm}
\usepackage{booktabs}
\usepackage{cite}
\usepackage{graphicx}
\usepackage{algorithm}
\usepackage{algorithmic}
\usepackage{mathtools}
\usepackage{url}
\usepackage{color}
\usepackage{placeins}

\newcommand{\bw}{\mathbf{w}}

\newcommand{\Ai}{\operatorname{Ai}}

\begin{document}

\title{Physics-Guided Neural Airy Beamforming for Near-Field Blockage Mitigation}

	\author{Yi~Wang and Linglong~Dai, {\textit{Fellow, IEEE}}
	\thanks{This work was supported in part by the National Science Fund for Distinguished Young Scholars under Grant 62325106, in part by the National Science and Technology Major Projects of China under Grant 2025ZD1301800, and in part by the National Key Research and Development Program of China under Grant 2023YFB3811503.}
	\thanks{The authors are with the Department of Electronic Engineering, Tsinghua University, and the State Key Laboratory of Space Network and Communications, Tsinghua University, Beijing 100084, China (e-mails: yiwang24@mails.tsinghua.edu.cn, daill@tsinghua.edu.cn).}
}

\maketitle

\begin{abstract}
High-frequency communication systems heavily rely on line-of-sight (LoS) paths, so blockage of the LoS path can cause severe performance loss. Near-field Airy beams with curved trajectories can steer energy around obstacles, offering a promising solution for blockage mitigation. However, existing methods for selecting a near-optimal Airy beam trajectory either rely on high-overhead beam training, or employ data-driven learning without a clear, physically interpretable rule.
To address this problem, we propose a physics-guided neural Airy beamforming framework that selects a near-optimal trajectory in one shot with clear physical interpretability.
Specifically, we first formulate a single-edge representation of the blocker model in 3GPP TR 38.901 and reveal the trajectory--edge coupling mechanism. This analysis yields a trajectory-selection optimality condition that defines the candidate trajectories. Although these trajectories generally cannot be expressed in closed form, we show that they form a continuous structure. This continuous structure is then exploited to construct a compact physics-defined region that captures near-optimal trajectories.
Guided by this region, a lightweight neural predictor is finally designed to directly select a near-optimal trajectory without beam training.
Simulations show that the compact physics-defined region effectively captures near-optimal trajectories, while occupying only about \(6\%\) of the candidate-space area on average. The proposed framework retains \(99.7\%\) of the reference rate obtained through numerical optimization, and nearly matches the rate of the data-driven method despite using approximately \(112\times\) fewer neural-network parameters.
\end{abstract}


\section{Introduction}\label{sec:introduction}

The growing demand for high data rates is driving communication systems toward the millimeter-wave and Terahertz bands~\cite{2021THz}. At these high frequencies, non-line-of-sight (NLoS) paths are typically much weaker than the line-of-sight (LoS) path, making reliable links strongly dependent on propagation along the LoS path~\cite{XLMIMO2}. Consequently, blockage of the LoS path can cause severe degradation in received power and thus achievable rate~\cite{galeote2026blockage}. Near-field Airy beams, with curved trajectories, can steer energy around obstacles toward the target receiver, offering a promising approach to mitigate blockage~\cite{darsena2025airy}.

\subsection{Prior Works}

Airy beams were originally predicted as nonspreading wave packets in quantum physics and later demonstrated in optics~\cite{airy1,airyfinite,airy5}. The special properties of Airy beams — self-acceleration, approximate nondiffracting propagation, and self-healing — have since been extensively studied~\cite{overview,selfhealing1,selfhealing2}. Specifically, the Airy beam follows a curved trajectory (self-acceleration), approximately maintains a fixed structure over distance (nondiffracting propagation), and recovers its shape beyond the obstacle (self-healing). Inspired by the blockage-robustness properties demonstrated in optics, recent studies have adapted the Airy-beam concept from optics to wireless communications for blockage mitigation. For example, Airy beams have been evaluated in near-field blockage scenarios, demonstrating improved blockage robustness over conventional focused beams~\cite{hanchong,songlingyang,wang2026multiairy}. To generate Airy beams in practical systems, array-based generation methods have been analyzed without using optical lenses~\cite{hanchong2}. Furthermore, experimental validations based on hardware platforms have confirmed the feasibility of Airy beams for wireless communications~\cite{lee,curving}.

A fundamental issue for Airy beams is how to select a near-optimal trajectory from a vast continuous candidate space. Existing methods address this issue through two main approaches: search-based beam training and learning-based direct prediction. Specifically, the search-based beam training~\cite{zhao2026efficienttraining} converts the continuous trajectory-selection problem into a discrete search problem over a large set of candidate beams. This simple yet effective scheme is able to achieve reliable performance. Nevertheless, each candidate evaluation requires receiver feedback, so the total overhead scales with the number of candidates, which may fail to meet real-time requirements. By contrast, the learning-based prediction scheme~\cite{chenhz} learns a direct mapping from the blockage geometry to Airy-beam control parameters, thereby avoiding the high-overhead search. However, this one-shot mapping treats trajectory selection as a black-box mapping process, and thus the physically interpretable rule for selecting a near-optimal Airy trajectory under blockage remains unclear.

\subsection{Our Contributions}

To establish a clear, physically interpretable rule for selecting a near-optimal Airy beam trajectory, we propose a physics-guided neural Airy beamforming framework that selects a near-optimal trajectory in one shot.\footnote{Simulation codes will be provided to reproduce the results in this article: \url{http://oa.ee.tsinghua.edu.cn/dailinglong/publications/publications.html}.} Specifically, the main contributions of this paper are summarized as follows.
\begin{itemize}
	\item First, we formulate a single-edge representation of the blocker model in 3GPP TR 38.901 and reveal the trajectory--edge coupling mechanism. This mechanism shows that, under blockage, the received power is jointly determined by an unblocked contribution and an edge-diffracted contribution. When the trajectory stays outside the obstacle edge, the unblocked contribution dominates. As the trajectory gradually moves toward the obstacle, the edge-diffracted contribution increases while the unblocked contribution decreases. This tradeoff gives rise to a received-power maximum and thus provides the physical basis to derive an optimality condition for Airy-beam trajectory selection.
	\item Based on this optimality condition, we apply the implicit function theorem~\cite{rudin1976principles} to characterize the associated trajectory structure of Airy beams. In particular, trajectories satisfying the optimality condition are not seperate solutions, but they form a continuous structure. This continuous structure can be exploited to construct a compact physics-defined region. The region captures near-optimal trajectories and significantly reduces the range of trajectories that should be considered. For a given blockage geometry, however, locating the optimal trajectory within this region still requires numerical optimization.
	\item To avoid the per-geometry numerical optimization, we train a lightweight neural predictor that directly maps the blockage geometry to a near-optimal Airy beam trajectory. The proposed predictor thus preserves the physical interpretability of trajectory selection while replacing per-geometry numerical optimization with a single predictor evaluation.
	\item Simulations show that the compact physics-defined region effectively captures near-optimal trajectories, while occupying only about $6\%$ of the full candidate-space area on average. Overall, the proposed framework retains $99.7\%$ of the rate obtained through numerical optimization. Compared with conventional Airy beam-training methods, the proposed framework achieves higher rates and reduces the beam-training overhead to one transmission. Moreover, the proposed physics-guided neural Airy beamforming also achieves performance comparable to the data-driven method, while requiring approximately $112\times$ fewer neural-network parameters.
\end{itemize}

\subsection{Organization and Notation}

\subsubsection{Organization}
The remainder of this paper is organized as follows.
Section~\ref{sec:system} introduces the system model and the Airy beam model.
Section~\ref{sec:edge_theory} develops the trajectory--edge coupling mechanism and the corresponding compact trajectory region.
Section~\ref{sec:selector} presents the physics-guided one-shot predictor, the online procedure, and the complexity analysis.
Section~\ref{sec:simulations} presents simulation results that validate the proposed method and compare it with baseline schemes.
Section~\ref{sec:conclusion} concludes the paper.

\subsubsection{Notation}
${\bf a}^H$ and ${\bf A}^H$ denote the conjugate transpose of vector ${\bf a}$ and matrix ${\bf A}$, respectively. The notation $\|{\bf a}\|_2$ denotes the Euclidean norm of ${\bf a}$, \(\operatorname{supp}(\cdot)\) denotes the support of a function, and \(\mathbb{1}(\cdot)\) denotes the indicator function. The imaginary unit is \(j=\sqrt{-1}\), and \(|\mathcal X|\) denotes the length of an interval \(\mathcal X\). The operator \(\mathcal{O}(\cdot)\) is used for complexity order.

\section{System Model and Airy Beam}\label{sec:system}

We first specify the near-field array-to-receiver geometry and the corresponding discrete channel response, and then introduce the single-edge blockage model. The Airy beam generation model is subsequently presented to formulate trajectory selection.

\subsection{Near-Field Channel Model}

We consider a narrowband near-field communication system in the $x$-$z$ plane, where an $N$-element uniform linear array (ULA) transmitter is centered at the origin and placed along the $x$-axis at $z=0$. The single-user receiver is represented by a finite effective receiving aperture centered at $(z_r,x_r)$.
The coordinate of the $n$-th antenna is
\begin{equation}
	x_n = \left(n-\frac{N+1}{2}\right)d,\quad n=1,\ldots,N,
\end{equation}
where $d\leq \lambda/2$ is the antenna spacing, $\lambda=c/f_c$ is the wavelength, $c$ is the speed of light, and $f_c$ is the carrier frequency. The aperture width and transverse aperture interval are $D=(N-1)d$ and $\mathcal X_{\rm tx}=[-D/2,D/2]$, respectively.
Throughout this paper, a position is written as $(z,x)$, where $z$ is the propagation distance and $x$ is the transverse coordinate.
We focus on the radiating near-field regime, where the receiver range is on the order of or below the Rayleigh distance $2D^2/\lambda$. For the unblocked reference link, the LoS channel vector $\mathbf h_{\rm LoS}\in\mathbb C^{N\times1}$ from the ULA to the receiver center is modeled as
\begin{equation}
	[\mathbf h_{\rm LoS}]_n
	=
\frac{\alpha_0}{L_n}\exp(jkL_n),
	\quad n=1,\ldots,N,
	\label{eq:near_field_channel}
\end{equation}
where $\alpha_0$ collects the distance-independent propagation and antenna constants, $k=2\pi/\lambda$ is the wavenumber, and
\begin{equation}
	L_n=\sqrt{z_r^2+(x_r-x_n)^2}
	\label{eq:element_receiver_distance}
\end{equation}
is the distance from the $n$-th antenna to the receiver center. Let $r=\sqrt{z_r^2+x_r^2}$ and $\sin\theta_r=x_r/r$. Expanding $L_n$ around the array center gives
\begin{equation}
	L_n
	\approx
	r-x_n\sin\theta_r
	+\frac{x_n^2\cos^2\theta_r}{2r}.
	\label{eq:near_field_distance_expansion}
\end{equation}
The linear term describes angular steering, whereas the range-dependent quadratic term captures near-field focusing and is absent from the far-field steering-vector approximation. Under the paraxial condition, \eqref{eq:near_field_distance_expansion} leads to the Fresnel propagation model used below. We use its continuous-aperture form to analyze the blocked channel and sample the resulting aperture excitation at $\{x_n\}_{n=1}^{N}$ for discrete ULA implementation. The center-point channel in \eqref{eq:near_field_channel} establishes the spherical-wave phase variation, while received power is evaluated by averaging the continuous field over the finite receiver window introduced below.

\subsection{Obstacle-Induced Blockage Model}

\begin{figure}[!t]
	\centering
	\includegraphics[width=\columnwidth]{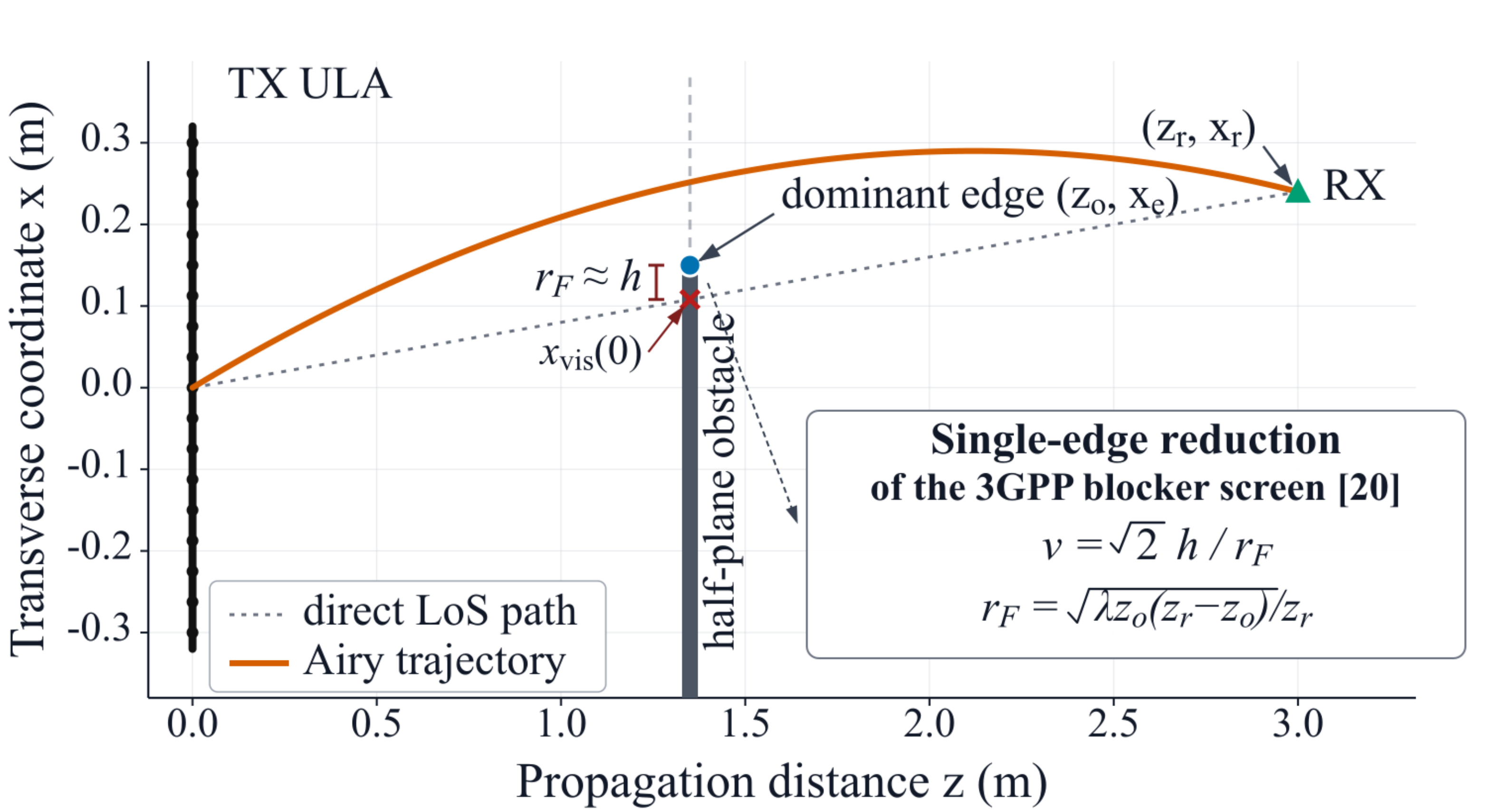}
	\caption{Near-field Airy beamforming geometry under the zero-thickness single-edge blockage model.}
	\label{fig:system_model}
\end{figure}

As shown in Fig.~\ref{fig:system_model}, a transient opaque obstacle lies between the transmitter and the receiver. At high frequencies, penetration and diffuse scattering through the obstacle can be much weaker than the surviving diffracted field, so we focus on the field around the relevant obstacle boundary.
The relevant obstacle boundary is represented by an edge point $(z_o,x_e)$ in the $x$-$z$ plane.
The unobstructed side of the edge is identified by $s\in\{+1,-1\}$, where
\begin{equation}
	s(x-x_e)>0
\end{equation}
represents the unobstructed half-plane, while $s(x-x_e)\leq0$ is blocked by the screen.
In 3GPP TR~38.901~\cite{3gpp38901}, each blocker is represented by a screen whose attenuation is computed from knife-edge diffraction at four edges. We retain the locally dominant edge and approximate the blocked side by a zero-thickness opaque half-plane at $z=z_o$. For the center aperture point, \(x_{\rm vis}(0)=z_o x_r/z_r\) is where its direct LoS path crosses the blockage plane, \(h=|x_e-x_{\rm vis}(0)|\) is the edge clearance, and \(r_F=\sqrt{\lambda z_o(z_r-z_o)/z_r}\) is the corresponding Fresnel radius. The corresponding normalized-clearance magnitude \(|\nu|=\sqrt{2}h/r_F\) is consistent with the Fresnel scaling in Recommendation ITU-R P.526~\cite{iturp526}. This single-edge representation also provides a local approximation to a larger obstacle boundary when one edge dominates the diffracted field.

Consistent with \eqref{eq:near_field_distance_expansion}, let $\mathcal{D}_{z}$ denote the scalar Fresnel free-space propagation operator over distance $z$.
The single-edge blockage mask at the blockage plane is
\begin{equation}
	M_o(x;x_e,s)=\mathbb{1}\{s(x-x_e)>0\}.
	\label{eq:blockage_mask}
\end{equation}
For an aperture excitation $u_0(x)$, whose Airy form is specified below, the blocked field at the receiver plane is modeled as
\begin{equation}
	U_{\rm blk}(x,z_r)
	=
	\mathcal{D}_{z_r-z_o}
	\left\{
	M_o(\cdot;x_e,s)
	\mathcal{D}_{z_o}\{u_0\}(\cdot)
	\right\}(x).
	\label{eq:blocked_propagation_operator}
\end{equation}
Arguments fixed by the scene geometry and aperture excitation are suppressed unless needed explicitly.
The corresponding free-space field is $U_{\rm free}(x,z_r)=\mathcal{D}_{z_r}\{u_0\}(x)$. Under the knife-edge interpretation, $U_{\rm blk}$ accounts for both the unblocked contribution that propagates through the unobstructed half-plane and the edge-diffracted contribution. Equation~\eqref{eq:blocked_propagation_operator} evaluates their combined field directly under the scalar Fresnel model.

The blockage ratio $\rho$ is defined from the geometrically unobstructed part of the transmit aperture.
For an aperture point $x\in\mathcal{X}_{\mathrm{tx}}$, the ray from $x$ to the receiver crosses the blockage plane at $x_{\rm vis}(x)=x+z_o(x_r-x)/z_r$.
The geometrically unobstructed aperture set and blockage ratio are
\begin{equation}
	\mathcal{X}_{\mathrm{vis}}
	=
	\{x\in\mathcal{X}_{\mathrm{tx}}:s[x_{\rm vis}(x)-x_e]>0\},
	\quad
	\rho
	=
	1-\frac{|\mathcal{X}_{\mathrm{vis}}|}{|\mathcal{X}_{\mathrm{tx}}|}.
	\label{eq:blockage_ratio_definition}
\end{equation}

The transmitter knows the receiver position, the relevant obstacle edge, and the unobstructed-side sign from sensing, environment perception, or a higher-layer map. Accordingly, the analysis and method are developed for the sensed single-edge geometry described above. Under this geometry, $\rho$ in \eqref{eq:blockage_ratio_definition} is derived from these quantities and the known aperture rather than introduced as an additional sensing assumption. Complete post-blockage channel state information (CSI) and receiver-feedback candidate sweeping are not required.

\subsection{Airy Beam Trajectory Control}

\begin{figure}[!t]
	\centering
	\includegraphics[width=\columnwidth]{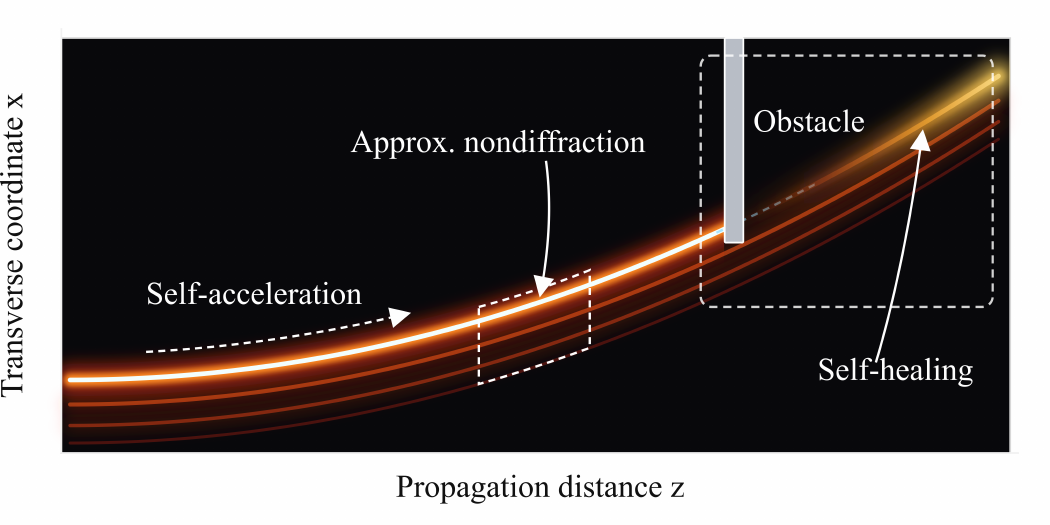}
	\caption{Finite-energy Airy-beam propagation, illustrating self-acceleration, approximate nondiffracting propagation, and self-healing after partial blockage.}
	\label{fig:airy_beam_properties}
\end{figure}

At the paraxial scale, a scalar field $\psi(x,z)$ satisfies
\begin{equation}
\left(\frac{\partial^2}{\partial x^2}+2jk\frac{\partial}{\partial z}\right)\psi(x,z)=0,
\qquad
\psi_{\rm id}(x,0)=\Ai(\gamma_{\rm A}x),
\label{eq:ideal_airy_boundary}
\end{equation}
where $\psi_{\rm id}$ denotes the ideal one-dimensional Airy field and $\gamma_{\rm A}$ is its transverse scaling parameter. The resulting free-space solution is~\cite{airy1,airy5}
\begin{equation}
\begin{aligned}
\psi_{\rm id}(x,z)
={}&\Ai\left(\gamma_{\rm A}x-
\frac{\gamma_{\rm A}^4z^2}{4k^2}\right)\\
&\times\exp\left\{j\frac{\gamma_{\rm A}^2z}{2k}
\left(\gamma_{\rm A}x-\frac{\gamma_{\rm A}^4z^2}{6k^2}\right)\right\}.
\end{aligned}
\label{eq:ideal_airy_solution}
\end{equation}
The Airy main-lobe peak therefore follows the parabolic trajectory $x_{\rm m}^{\rm id}(z)=\xi_{\rm p}/\gamma_{\rm A}+\gamma_{\rm A}^3z^2/(4k^2)$, where $\xi_{\rm p}$ denotes the location of the principal maximum of $\Ai(\cdot)$. This deterministic bending gives the self-acceleration property. The ideal solution preserves its transverse profile, while a finite-energy realization approximately retains this structure and can reconstruct a weakened main lobe from the remaining sidelobes after partial blockage~\cite{overview,selfhealing1,selfhealing2}.

Figure~\ref{fig:airy_beam_properties} summarizes these propagation properties. The curved dashed guide marks the parabolic trajectory, the narrow dashed contour highlights the nearly preserved main-lobe width before blockage, and the downstream dashed box encloses the interruption and gradual reconstruction caused by the knife edge.

Because the ideal Airy profile carries infinite energy, practical generation requires finite-energy apodization~\cite{airyfinite}. In the array-based realization adopted here, a finite Gaussian aperture is combined with cubic, quadratic, and linear phase terms to control bending, focusing, and steering, respectively.

Following the established near-field Airy generation method~\cite{hanchong2,zhao2026efficienttraining}, the continuous initial aperture field at transverse coordinate $x_0$ is written as
\begin{equation}
	u_0(x_0;B,F,\theta)
	=
	a_0(x_0)
\exp\{j\phi_{\mathrm A}(x_0;B,F,\theta)\},
	\label{eq:aperture_field}
\end{equation}
where $a_0(x_0)=\exp(-x_0^2/\omega_0^2)\mathbb{1}\{|x_0|\leq D/2\}$ is the finite Gaussian aperture envelope of waist $\omega_0$ and aperture width $D$.
The phase term in \eqref{eq:aperture_field} is
\begin{equation}
	\phi_{\mathrm A}(x_0;B,F,\theta)
	=
	\frac{1}{3}(2\pi B)^3x_0^3
	-\frac{\pi}{\lambda F}x_0^2
	+\frac{2\pi}{\lambda}\sin\theta\,x_0,
	\label{eq:airy_phase}
\end{equation}
where $B$ controls the cubic bending strength, while $F$ and $\theta$ specify the distance and angular direction of the Airy generation-plane center. Throughout this paper, $(B,F,\theta)$ is called the Airy control triplet. The implementable ULA excitation samples \eqref{eq:aperture_field} at the antenna positions and normalizes the transmit power:
\begin{equation}
\begin{aligned}
\tilde w_n&=u_0(x_n;B,F,\theta),\quad n=1,\ldots,N,\\
\tilde{\bw}&=[\tilde w_1,\ldots,\tilde w_N]^{\mathrm T},\qquad
\bw(B,F,\theta)=\frac{\tilde{\bw}}{\|\tilde{\bw}\|_2}.
\end{aligned}
\label{eq:discrete_airy_beamforming_vector}
\end{equation}
Here, $x_0$ is the continuous coordinate used in the Fresnel analysis, whereas $x_n$ indexes the physical ULA elements.

\begin{figure}[!t]
	\centering
	\includegraphics[width=\columnwidth]{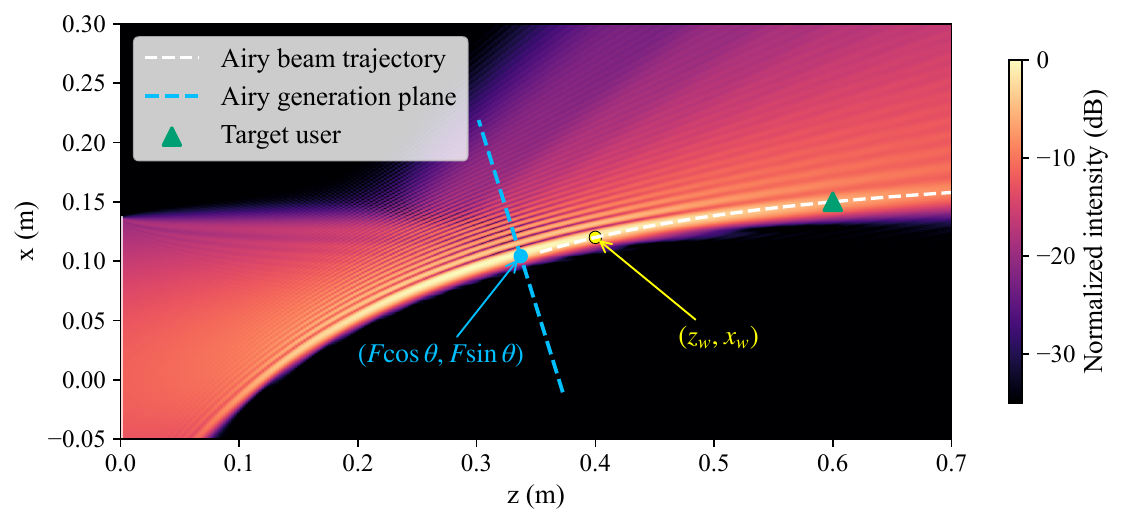}
	\caption{Airy-beam trajectory control through the Airy control triplet $(B,F,\theta)$ and waypoint $(z_w,x_w)$.}
	\label{fig:airy_trajectory_control}
\end{figure}

Under the Gaussian-aperture and Fresnel approximations, the Airy main-lobe trajectory can be expressed as~\cite{hanchong2,zhao2026efficienttraining}
\begin{equation}
	x_{\mathrm m}(z)
	=
	-\xi_{\mathrm p}\lambda B z
	+\sin\theta\,z
	-
	\frac{S_R^2(z)-S_I^2}{16\lambda\pi^2B^3}z,
	\label{eq:airy_trajectory}
\end{equation}
where $\xi_{\mathrm p}\approx -1.0188$ denotes the first maximum of the Airy function~\cite{airyfinite},
\begin{equation}
	S_R(z)=\frac{1}{z}-\frac{1}{F},\qquad
	S_I=\frac{\lambda}{\pi\omega_0^2}.
\end{equation}
Figure~\ref{fig:airy_trajectory_control} illustrates the resulting no-blockage trajectory, the Airy generation plane specified by $(F,\theta)$, and the trajectory waypoint $(z_w,x_w)$ used in the analytical generation map below.

Existing work provides a closed-form map from a waypoint $(z_w,x_w)$ and receiver position $(z_r,x_r)$ to an Airy control triplet:
\begin{equation}
	(B,F,\theta)=\mathcal{G}_{\mathrm A}(z_w,x_w;z_r,x_r,s).
	\label{eq:waypoint_to_params}
\end{equation}
Here, the unobstructed-side sign $s$ selects the corresponding bending branch. The exact algebraic form of $\mathcal{G}_{\mathrm A}(\cdot)$ is available in~\cite{hanchong2,zhao2026efficienttraining}.
In this paper, \eqref{eq:waypoint_to_params} is used as an established generation operator.

The ideal one-shot trajectory selection problem can now be stated as follows.
Given the receiver position, the relevant obstacle edge, and the unobstructed-side sign,
\begin{equation}
	\mathcal{I}=\{z_r,x_r,z_o,x_e,s\},
\end{equation}
let $\mathcal R=[x_r-W_r/2,x_r+W_r/2]$ denote the receiver window of width $W_r$. For an Airy control triplet \(p=(B,F,\theta)\), we make the field dependence explicit as \(U_{\rm blk}(x,z_r;p,\mathcal I)\) and define
\begin{equation}
P_{\rm rx}(p;\mathcal I)=\frac{1}{W_r}\int_{\mathcal R}|U_{\rm blk}(x,z_r;p,\mathcal I)|^2\,\mathrm{d}x.
\label{eq:rx_power}
\end{equation}
For fixed transmit power and noise power, maximizing $P_{\rm rx}$ is equivalent to maximizing the received signal-to-noise ratio (SNR) and achievable rate. The transmitter therefore selects \(p\) without receiver-feedback candidate sweeping:
\begin{equation}
	p^\star
	=
	\arg\max_{p\in\mathcal{P}_{\mathrm A}}
	P_{\rm rx}(p;\mathcal{I}),
	\label{eq:continuous_airy_selection_problem}
\end{equation}
Here, \(p\) generates the Airy beamforming vector \(\bw(p)\) through \eqref{eq:discrete_airy_beamforming_vector}, and \(\mathcal{P}_{\mathrm A}\) is the physically feasible set of Airy control triplets under the adopted aperture and paraxial-propagation assumptions.
Under blockage, \eqref{eq:continuous_airy_selection_problem} has no tractable closed-form maximizer because diffraction, sidelobes, finite aperture, and receiver-window averaging are coupled.
Through the established map \eqref{eq:waypoint_to_params}, the remaining task is therefore to identify a near-optimal waypoint under blockage. The next section develops the physical structure needed for this task.

\section{Trajectory--Edge Coupling Mechanism}\label{sec:edge_theory}

This section establishes the physical basis of the proposed method. We first derive the finite-aperture single-edge Fresnel response and use the balance between the unblocked and edge-diffracted contributions to reveal the trajectory--edge coupling mechanism. We then express the Airy trajectory in Fresnel-scaled coordinates and derive the exact received-power derivatives and the resulting optimality condition. Finally, the implicit function theorem and local power curvature are used to characterize the continuous trajectory structure and construct the compact physics-defined region.

\subsection{Single-Edge Fresnel Response}

Under the scalar paraxial approximation, we use the standard one-dimensional Fresnel kernel~\cite{optics}, where $k=2\pi/\lambda$ is the wavenumber. Let $L_1=z_o$ and $L_2=z_r-z_o$. The variables $x_a$ and $x_b$ denote the input and output transverse coordinates:
\begin{equation}
K_L(x_b,x_a)=\frac{e^{jkL}}{\sqrt{j\lambda L}}
\exp\left\{\frac{j\pi}{\lambda L}(x_b-x_a)^2\right\}.
\label{eq:fresnel_kernel}
\end{equation}
For a finite aperture $x_0\in[-D/2,D/2]$, the field under blockage at an output coordinate $x_b$ is
\begin{equation}
\begin{aligned}
U_{\rm blk}(x_b)=\int_{-D/2}^{D/2}u_0(x_0)
&\int_{s(x-x_e)>0}K_{L_2}(x_b,x)\\
&\quad\times K_{L_1}(x,x_0)\,\mathrm{d}x\,\mathrm{d}x_0.
\end{aligned}
\label{eq:blocked_two_hop_field}
\end{equation}
Completing the square in the inner integral gives
\begin{equation}
\frac{(x-x_0)^2}{L_1}+\frac{(x_b-x)^2}{L_2}
=\frac{(x_b-x_0)^2}{z_r}+\frac{(x-x_c(x_0;x_b))^2}{L_{\rm eff}},
\label{eq:fresnel_square_completion}
\end{equation}
where
\begin{equation}
x_c(x_0;x_b)=\frac{L_2x_0+L_1x_b}{z_r},\qquad
L_{\rm eff}=\frac{L_1L_2}{z_r}.
\label{eq:effective_fresnel_distance}
\end{equation}
The obstacle-plane half-line in \eqref{eq:blocked_two_hop_field} therefore has the following analytical primitive:
\begin{equation}
\begin{aligned}
&\int_{s(x-x_e)>0}K_{L_2}(x_b,x)K_{L_1}(x,x_0)\,\mathrm{d}x\\
&\qquad=K_{z_r}(x_b,x_0)H_s(x_0;x_b),
\end{aligned}
\label{eq:halfline_fresnel_primitive}
\end{equation}
with
\begin{equation}
H_s(x_0;x_b)=\frac{1}{2}\operatorname{erfc}\!\left[
s e^{-j\pi/4}
\sqrt{\frac{\pi}{\lambda L_{\rm eff}}}
\bigl(x_e-x_c(x_0;x_b)\bigr)\right].
\label{eq:edge_transfer_function}
\end{equation}
Here, $x_c(x_0;x_b)$ is the obstacle-plane center of the composed Fresnel phase and $\operatorname{erfc}(\cdot)$ is the complementary error function.
For the Gaussian-cubic aperture field in \eqref{eq:aperture_field}--\eqref{eq:airy_phase}, this produces a single finite integral rather than a two-step propagation simulation:
\begin{equation}
\begin{aligned}
U_{\rm blk}(x_b)&\propto\int_{-D/2}^{D/2}
e^{j\Psi(x_0;x_b,z_r)}H_s(x_0;x_b)\,\mathrm{d}x_0,\\
\Psi(x_0;x_b,z_r)&=\frac{(2\pi B)^3}{3}x_0^3
+C_2(z_r)x_0^2+C_1(x_b,z_r)x_0.
\end{aligned}
\label{eq:composed_blocked_field}
\end{equation}
where the coefficient functions are
\begin{equation}
\begin{aligned}
C_1(x,z)&=\frac{2\pi}{\lambda}\left(\sin\theta-\frac{x}{z}\right),\\
C_2(z)&=\frac{\pi}{\lambda}\left(\frac{1}{z}-\frac{1}{F}\right)+\frac{j}{\omega_0^2}.
\end{aligned}
\label{eq:composed_phase_coefficients}
\end{equation}
The imaginary component of $C_2(z)$ retains the Gaussian aperture envelope. The proportionality in \eqref{eq:composed_blocked_field} omits only a factor independent of $(B,F,\theta)$. This factor is retained in the absolute-power and rate evaluations. Writing $H_s=1+(H_s-1)$ separates the blocked field into the unblocked field and an edge-diffracted correction, which are the two physical contributions examined below. The expression jointly represents finite aperture, Gaussian-cubic Airy generation, single-edge diffraction, and receiver coupling under the stated scalar Fresnel model.

\subsection{Edge-Induced Power Variation}

The Fresnel kernel and the single-edge response above are standard propagation ingredients~\cite{optics}. We next isolate how displacement of the edge relative to a fixed trajectory changes the balance between the unblocked and edge-diffracted contributions.

Under the single-edge model, the edge dependence of the metric in \eqref{eq:rx_power} is exact for any continuous field incident on the obstacle plane. The receiver window $\mathcal R$ is fixed by the receiver position and does not vary with $x_e$. Let $U_o(x)=\mathcal D_{z_o}\{u_0\}(x)$ denote the transmitter-excited field at the obstacle plane before masking. Applying Leibniz's rule to the moving boundary of the half-line $s(x-x_e)>0$ gives, for any receiver-plane coordinate $x_b$,
\begin{equation}
\frac{\partial U_{\rm blk}(x_b;x_e)}{\partial x_e}
=-sK_{L_2}(x_b,x_e)U_o(x_e).
\label{eq:edge_field_derivative}
\end{equation}
Consequently,
\begin{equation}
\frac{\partial P_{\rm rx}}{\partial x_e}
=-\frac{2s}{W_r}\operatorname{Re}\!\left\{
\int_{\mathcal R}U_{\rm blk}^*(x;x_e)K_{L_2}(x,x_e)U_o(x_e)\,\mathrm{d}x
\right\}.
\label{eq:edge_power_derivative}
\end{equation}
Equation~\eqref{eq:edge_power_derivative} makes the trajectory--edge coupling mechanism precise. For a fixed Airy trajectory, an infinitesimal edge displacement adds or removes the boundary contribution $K_{L_2}(x,x_e)U_o(x_e)$. The derivative therefore quantifies the received-power sensitivity to the relative edge position. When the trajectory passes well into the unobstructed half-plane, the unblocked contribution dominates and edge illumination is weak. Moving the trajectory toward the edge strengthens the edge-diffracted contribution. Moving too far toward the blocked half-plane increasingly weakens the unblocked contribution. Along a one-dimensional edge-displacement sweep, a local balance between the two contributions satisfies
\begin{equation}
\frac{\partial P_{\rm rx}}{\partial x_e}=0,
\qquad
\frac{\partial^2 P_{\rm rx}}{\partial x_e^2}<0.
\label{eq:edge_coupling_stationarity}
\end{equation}

\begin{figure*}[!t]
	\centering
	\includegraphics[width=0.98\textwidth]{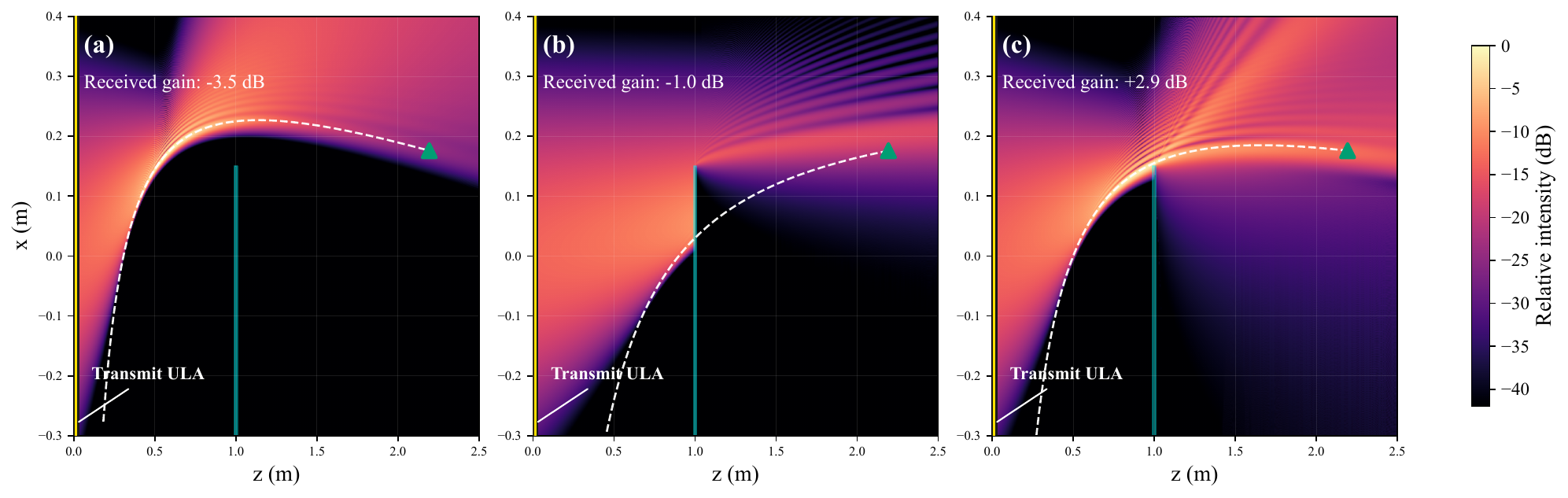}
	\caption{Full-space intensity distributions for representative trajectory--edge configurations: (a) unblocked-contribution dominance; (b) edge-diffraction dominance; and (c) coupled optimum.}
	\label{fig:edge_interaction_mechanism}
\end{figure*}

Figure~\ref{fig:edge_interaction_mechanism} contrasts the two limiting regimes with the received-power maximum between them. The dashed curves mark the corresponding unblocked Airy trajectories, the highlighted rectangles at $z=0$ mark the transmit ULA, and the annotated gains are measured relative to the focused free-space benchmark.
In Fig.~\ref{fig:edge_interaction_mechanism}(a), the trajectory passes outside the obstacle edge, and the unblocked contribution dominates. In Fig.~\ref{fig:edge_interaction_mechanism}(b), the trajectory is directed toward the blocked half-plane, and the receiver relies primarily on edge diffraction after the unblocked contribution is weakened. The intermediate trajectory in Fig.~\ref{fig:edge_interaction_mechanism}(c) produces the most favorable balance between the two contributions. The three regimes illustrate the trajectory--edge coupling mechanism and motivate an edge-scaled trajectory representation. Equation~\eqref{eq:edge_coupling_stationarity} characterizes the edge-displacement response for a fixed trajectory. Because trajectory adjustment jointly changes all three control parameters, the trajectory-selection optimality condition is derived from the exact trajectory derivatives below.

\subsection{Edge-Conditioned Trajectory Coordinates}

The transition factor in \eqref{eq:edge_transfer_function} varies when the composed Fresnel center is within an order of $\sqrt{\lambda L_{\rm eff}}$ of the obstacle edge. This motivates the transverse Fresnel scale
\begin{equation}
r_F=\sqrt{\frac{\lambda z_o(z_r-z_o)}{z_r}}=\sqrt{\lambda L_{\rm eff}},
\label{eq:fresnel_scale}
\end{equation}
and the normalized trajectory coordinates
\begin{equation}
z_w=z_o+\beta(z_r-z_o),\qquad
x_w=x_e+s\eta_w r_F.
\label{eq:waypoint_coordinates}
\end{equation}
The coordinate $\eta_w$ is the normalized transverse displacement of the waypoint from the obstacle edge, with $\eta_w>0$ on the unobstructed side and $\eta_w<0$ on the blocked side. The coordinate $0\leq\beta<1$ locates the waypoint between the obstacle and receiver planes. The deterministic composition
\begin{equation}
\mathbf q=(\eta_w,\beta)
\mapsto(z_w,x_w)
\mapsto(B,F,\theta)
\label{eq:trajectory_map}
\end{equation}
is obtained from \eqref{eq:waypoint_coordinates} and \eqref{eq:waypoint_to_params}. Thus, $\mathbf q$ jointly changes the cubic, focusing, and steering phases.

\subsection{Compact Trajectory Region}

Let $\varphi(x_0;\mathbf q)=\phi_{\rm A}[x_0;B(\mathbf q),F(\mathbf q),\theta(\mathbf q)]$. Since the aperture envelope and edge-transfer factor do not depend on $\mathbf q$, differentiating \eqref{eq:composed_blocked_field} under the finite integral gives, for $q\in\{\eta_w,\beta\}$,
\begin{equation}
\begin{aligned}
\frac{\partial U_{\rm blk}(x;\mathbf q)}{\partial q}
={}&j\int_{-D/2}^{D/2}
\varphi_q(x_0;\mathbf q)u_0(x_0;\mathbf q)\\
&\quad{}\times K_{z_r}(x,x_0)H_s(x_0;x)\,\mathrm{d}x_0,
\end{aligned}
\label{eq:trajectory_field_derivative}
\end{equation}
where the exact source-phase derivative is
\begin{equation}
\begin{aligned}
\varphi_q(x_0;\mathbf q)=
&(2\pi)^3B^2B_qx_0^3
+\frac{\pi F_q}{\lambda F^2}x_0^2\\
&+\frac{2\pi\cos\theta}{\lambda}\theta_qx_0.
\end{aligned}
\label{eq:source_phase_derivative}
\end{equation}
The quantities $(B_q,F_q,\theta_q)$ are obtained by differentiating the analytical map in \eqref{eq:trajectory_map}. Writing $P_q\triangleq\partial P_{\rm rx}/\partial q$, the corresponding received-power derivative is
\begin{equation}
P_q=\frac{2}{W_r}\operatorname{Re}\!\left\{
\int_{\mathcal R}U_{\rm blk}^*(x;\mathbf q)
\frac{\partial U_{\rm blk}(x;\mathbf q)}{\partial q}\,\mathrm{d}x
\right\}.
\label{eq:trajectory_power_derivative}
\end{equation}
Equation~\eqref{eq:trajectory_power_derivative} retains the joint variation of all three Airy controls, finite-aperture coupling, single-edge diffraction, and receiver-window averaging.

The relation to the trajectory--edge coupling mechanism is explicit. Let $P_{\rm free}(\mathbf q)$ denote the receiver-window power obtained from $U_{\rm free}$, and define the edge-induced power correction by $\Delta P_{\rm e}(\mathbf q)=P_{\rm rx}(\mathbf q)-P_{\rm free}(\mathbf q)$. At an interior optimum, the trajectory-selection condition is
\begin{equation}
\nabla_{\mathbf q}P_{\rm rx}=\mathbf 0
\quad\Longleftrightarrow\quad
\nabla_{\mathbf q}P_{\rm free}
=-\nabla_{\mathbf q}\Delta P_{\rm e}.
\label{eq:stationary_gradient_balance}
\end{equation}
Thus, the free-space preference and the edge-induced correction balance at an interior optimum. This gradient balance links the trajectory--edge coupling mechanism to the trajectory-selection optimality condition. In the fully unblocked limit, $H_s\rightarrow1$, $\rho\rightarrow0$, and $\nabla_{\mathbf q}\Delta P_{\rm e}\rightarrow\mathbf0$, so the condition reduces continuously to free-space trajectory selection.

Let $L_{\rm A}=z_r\cos\theta+x_r\sin\theta-F$ denote the remaining axial distance from the Airy generation-plane center to the receiver, and let $\varepsilon_{\max}(\mathbf q;\mathcal I)$ be the Fresnel phase remainder defined in Appendix~\ref{app:fresnel_feasibility}. For an allowed remainder $\tau_{\rm F}$, the analytically admissible trajectory set is
\begin{equation}
\begin{aligned}
\mathcal F_{\rm A}(\mathcal I)=\{\mathbf q:\;&
F>0,\ L_{\rm A}>0,\ |\sin\theta|<1,\\
&\varepsilon_{\max}(\mathbf q;\mathcal I)\leq\tau_{\rm F}\}.
\end{aligned}
\label{eq:feasible_trajectory_set}
\end{equation}
Here, the strict steering condition $|\sin\theta|<1$ keeps the inverse-sine branch in $\mathcal G_{\rm A}$ regular. To obtain a continuous reduced candidate structure, we impose transverse stationarity for each $\beta$ and retain $\beta$ for the final power maximization. Across $0\leq\beta<1$, define the transverse local-maximum set
\begin{equation}
\mathcal C(\mathcal I)=\{(\eta_w,\beta)\in\operatorname{int}\mathcal F_{\rm A}:
P_{\eta_w}=0,\ P_{\eta_w\eta_w}<0\}.
\label{eq:transverse_stationary_set}
\end{equation}
It may contain several branches because the finite Airy field has multiple lobes and the receiver power is nonconvex.

\noindent\textbf{Proposition 1 (local stationary branch and power-loss width).}
Suppose that $P_{\rm rx}>0$ is three times continuously differentiable near a nondegenerate transverse local maximum $(\eta_j,\beta_0)\in\mathcal C(\mathcal I)$, where $P_{\eta_w\eta_w}(\eta_j,\beta_0)<0$. Then a unique differentiable stationary branch $\eta_j(\beta)$ exists locally and satisfies
\begin{equation}
\frac{\mathrm d\eta_j}{\mathrm d\beta}
=-\frac{P_{\eta_w\beta}}{P_{\eta_w\eta_w}}.
\label{eq:stationary_branch_slope}
\end{equation}
Let $\kappa_j=-\partial^2\ln P_{\rm rx}/\partial\eta_w^2>0$ on this branch. For a transverse displacement $\delta_\eta$, the local power loss is
\begin{equation}
10\log_{10}\frac{P_{\rm rx}[\eta_j(\beta),\beta]}
{P_{\rm rx}[\eta_j(\beta)+\delta_\eta,\beta]}
=\frac{5\kappa_j}{\ln 10}\delta_\eta^2
+\mathcal O(|\delta_\eta|^3).
\label{eq:local_power_loss}
\end{equation}
Therefore, the second-order half-width associated with an $\epsilon_{\rm dB}$ local loss is
\begin{equation}
w_j(\beta;\epsilon_{\rm dB})
=\sqrt{\frac{\epsilon_{\rm dB}\ln 10}{5\kappa_j(\beta)}}.
\label{eq:curvature_width}
\end{equation}
Here, $w_j$ is measured in the normalized $\eta_w$ coordinate, and $r_Fw_j$ is the corresponding physical transverse half-width. Within $\operatorname{int}\mathcal F_{\rm A}$, the generation map remains on its regular branch and the finite-aperture Fresnel integrals are smooth in $\mathbf q$, so the assumed $C^3$ regularity holds locally.
The proof is given in Appendix~\ref{app:proof_stationary_branch}.

Proposition~1 establishes the continuous stationary structure underlying the trajectory candidates. This structure can include weak local maxima associated with finite-aperture sidelobes. Such a branch cannot determine the highest-power trajectory when a stronger stationary branch exists at the same $\beta$. To reduce the candidate structure further without discarding a potential maximizer, we retain only stationary branches that remain competitive in received power. For each $\beta$ at which $\mathcal C(\mathcal I)$ is nonempty, define the strongest transverse stationary power as
\begin{equation}
P_{\perp}^{\max}(\beta;\mathcal I)
=\max_{\eta_w:(\eta_w,\beta)\in\mathcal C(\mathcal I)}
P_{\rm rx}(\eta_w,\beta;\mathcal I).
\label{eq:strongest_transverse_power}
\end{equation}
Using the same received-power tolerance as in \eqref{eq:curvature_width}, the competitive stationary set is
\begin{equation}
\begin{aligned}
&\mathcal C_{\rm cmp}(\mathcal I;\epsilon_{\rm dB})\\[-0.2ex]
&\quad=\bigl\{(\eta_j(\beta),\beta)\in\mathcal C(\mathcal I):\\[-0.2ex]
&\qquad 10\log_{10}
\frac{P_{\perp}^{\max}(\beta;\mathcal I)}
{P_{\rm rx}[\eta_j(\beta),\beta;\mathcal I]}\\
&\qquad\leq\epsilon_{\rm dB}\bigr\}.
\end{aligned}
\label{eq:competitive_stationary_set}
\end{equation}
Any regular interior maximizer must attain the strongest transverse stationary power at its own $\beta$ and hence belongs to $\mathcal C_{\rm cmp}(\mathcal I;\epsilon_{\rm dB})$ with zero gap. The tolerance also retains nearly tied branches when their received-power ordering changes. Thus, \eqref{eq:competitive_stationary_set} removes weak stationary branches without excluding an interior power-maximizing trajectory.

\begin{figure}[!t]
\centering
\includegraphics[width=\columnwidth]{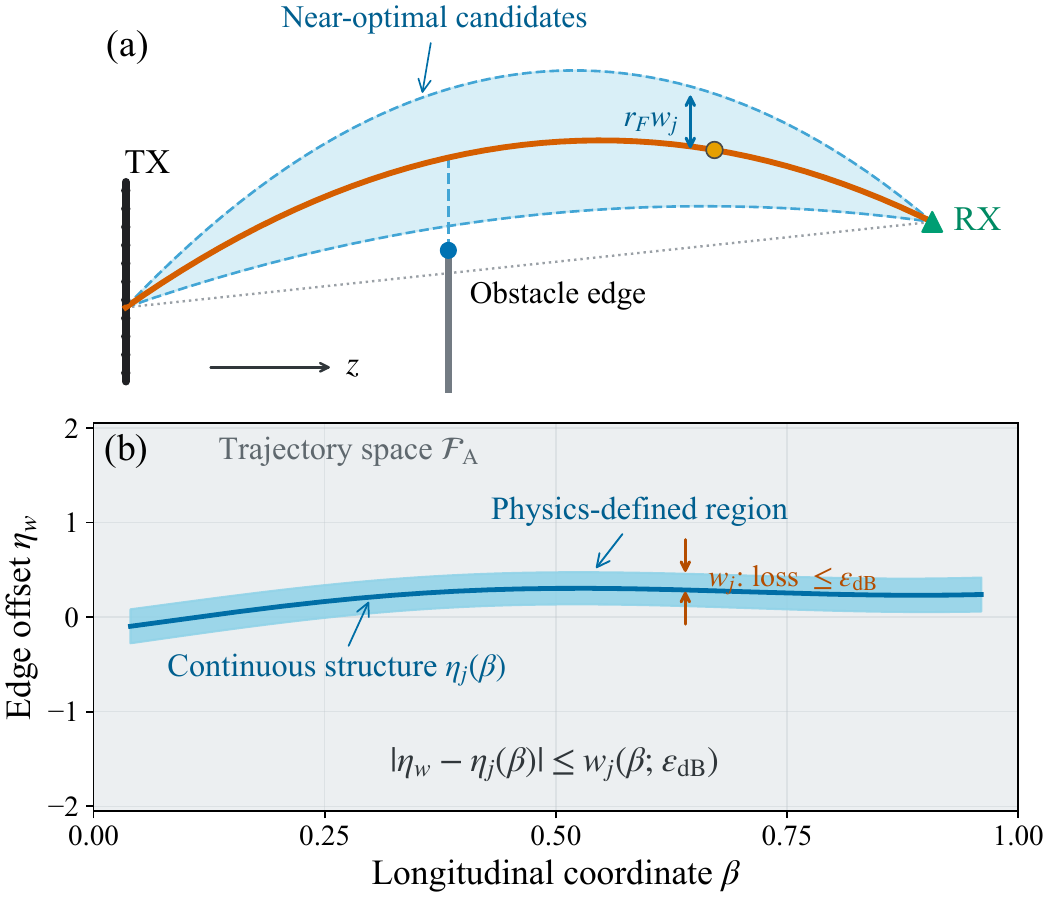}
\caption{Construction of the compact physics-defined trajectory region: (a) candidate trajectory neighborhood at fixed $\beta_0$; (b) continuous trajectory structure and the resulting region in the $(\beta,\eta_w)$ coordinate plane.}
\label{fig:waypoint_coordinates}
\end{figure}

Figure~\ref{fig:waypoint_coordinates} illustrates how the continuous trajectory structure is transformed into the physics-defined region after retaining the competitive stationary set in \eqref{eq:competitive_stationary_set}. At a fixed $\beta_0$, Fig.~\ref{fig:waypoint_coordinates}(a) maps the coordinate half-width $w_j(\beta_0;\epsilon_{\rm dB})$ to the physical transverse half-width $r_Fw_j(\beta_0;\epsilon_{\rm dB})$ and visualizes the resulting candidate trajectory neighborhood under the second-order local-loss approximation in \eqref{eq:local_power_loss}. Figure~\ref{fig:waypoint_coordinates}(b) shows how a retained branch of the continuous structure $\eta_j(\beta)$ is expanded by its curvature half-width within $\mathcal F_{\rm A}(\mathcal I)$ to form the physics-defined region.

The curvature neighborhoods of the competitive stationary branches define the interior branch region
\begin{equation}
\begin{aligned}
&\mathcal T_{\rm br}(\mathcal I;\epsilon_{\rm dB})\\[-0.2ex]
&\quad=\bigl\{(\eta_w,\beta)\in\mathcal F_{\rm A}(\mathcal I):\\[-0.2ex]
&\qquad\exists\,(\eta_j(\beta),\beta)
\in\mathcal C_{\rm cmp}(\mathcal I;\epsilon_{\rm dB}),\\[-0.2ex]
&\qquad|\eta_w-\eta_j(\beta)|
\leq w_j(\beta;\epsilon_{\rm dB})\bigr\}.
\end{aligned}
\label{eq:interior_branch_region}
\end{equation}
Let $\mathcal K_{\partial\mathcal F_{\rm A}}(\mathcal I)$ denote the regular candidates that satisfy the Karush--Kuhn--Tucker (KKT) condition on active boundaries of $\mathcal F_{\rm A}(\mathcal I)$. The complete physics-defined compact trajectory region is
\begin{equation}
\mathcal T_{\rm stat}(\mathcal I;\epsilon_{\rm dB})
=\mathcal T_{\rm br}(\mathcal I;\epsilon_{\rm dB})
\cup\mathcal K_{\partial\mathcal F_{\rm A}}(\mathcal I).
\label{eq:stationary_trajectory_region}
\end{equation}
Equation~\eqref{eq:stationary_trajectory_region} is scene dependent and may contain several competitive branches. It is not an empirically prescribed candidate grid. Every regular interior constrained maximizer belongs to \eqref{eq:competitive_stationary_set} and hence to \eqref{eq:interior_branch_region}, while a maximizer on an active boundary of $\mathcal F_{\rm A}(\mathcal I)$ is included through $\mathcal K_{\partial\mathcal F_{\rm A}}(\mathcal I)$. The compact trajectory region therefore removes weaker stationary branches while retaining the physics-defined candidates needed for trajectory selection.

\section{Physics-Guided Neural Airy Beamforming}\label{sec:selector}

Building on the compact physics-defined region in \eqref{eq:stationary_trajectory_region}, this section defines the stationary/KKT trajectory reference used for offline supervision. Maximizing received power over the stationary/KKT solution set yields a canonical reference trajectory. A lightweight neural predictor then learns the direct blockage-geometry-to-trajectory mapping so that online operation requires neither propagation integrals nor stationary-point calculations.

\subsection{Stationary/KKT Trajectory Reference}

Let $\mathcal Q$ be a finite numerical coordinate chart that contains the relevant Fresnel-scale stationary branches, and define $\Omega(\mathcal I)=\mathcal F_{\rm A}(\mathcal I)\cap\mathcal Q$. Let $\mathcal K_{\partial\Omega}(\mathcal I)$ contain the regular KKT points on active boundaries of $\mathcal F_{\rm A}(\mathcal I)$ or $\mathcal Q$. The stationary/KKT solution set over this chart is
\begin{equation}
\mathcal S(\mathcal I;\epsilon_{\rm dB})=
\left[\mathcal C_{\rm cmp}(\mathcal I;\epsilon_{\rm dB})
\cap\operatorname{int}\Omega(\mathcal I)\right]
\cup\mathcal K_{\partial\Omega}(\mathcal I).
\label{eq:stationary_kkt_set}
\end{equation}
The offline trajectory reference is
\begin{equation}
\mathbf q^\star(\mathcal I)=
\arg\max_{\mathbf q\in\mathcal S(\mathcal I;\epsilon_{\rm dB})}
P_{\rm rx}(\mathbf q;\mathcal I).
\label{eq:stationary_kkt_reference}
\end{equation}
All transverse roots in $\operatorname{int}\Omega(\mathcal I)$ are located from the analytical gradient in \eqref{eq:trajectory_power_derivative} and refined continuously along the branches within the chart. Their received powers determine the competitive stationary set in \eqref{eq:competitive_stationary_set}, while boundary candidates satisfy the projected KKT condition. Equation~\eqref{eq:stationary_kkt_reference} supplies the canonical supervision and evaluation target, with any exact tie resolved deterministically. Because $\mathbf q^\star$ already selects the highest-power member of the stationary/KKT solution set, the online predictor does not require an explicit branch index.

This construction is distinct from an online beam sweep. The broad waypoint reference used in Section~\ref{sec:simulations} expands the numerical chart and applies independent search initializations only to validate \eqref{eq:stationary_kkt_reference}. It is an offline comparison within the stated Airy generation method, not a global optimum over arbitrary array excitations.

\subsection{One-Shot Trajectory Predictor}

The neural input is
\begin{equation}
\mathbf g=[z_r,x_r,z_o,x_e,s,\rho]^{\mathrm T},
\label{eq:predictor_input}
\end{equation}
where the first five entries are sensed geometry quantities and $\rho$ is derived from them using \eqref{eq:blockage_ratio_definition}. Thus, the sensing requirement remains five-dimensional, while the predictor uses a six-dimensional feature vector. After standardizing the input with training-set statistics, a fully connected multilayer perceptron (MLP) with trainable weights $\bm\vartheta$ directly predicts the continuous trajectory coordinates,
\begin{equation}
(\hat\eta_w,\hat\beta)=f_{\bm\vartheta}(\mathbf g).
\label{eq:predictor_output}
\end{equation}
The two outputs are de-standardized with training statistics and bounded by the fixed numerical chart $\mathcal Q$ as a numerical safeguard. The bounded outputs are still denoted by $(\hat\eta_w,\hat\beta)$. Equations~\eqref{eq:waypoint_coordinates}, \eqref{eq:waypoint_to_params}, and \eqref{eq:discrete_airy_beamforming_vector} then generate one Airy beam without a candidate search.

The predictor is trained offline with scene-level splits. Its target is the stationary/KKT trajectory reference in \eqref{eq:stationary_kkt_reference}, and the loss is a normalized SmoothL1 coordinate loss. Checkpoints are selected by validation receiver-window power rather than coordinate error. The independent test set is used only for final evaluation. The predictor therefore learns the nonlinear blockage-geometry-to-trajectory map identified by the physical analysis. It does not evaluate propagation or run numerical optimization online.

\begin{figure}[!t]
\centering
\includegraphics[width=\columnwidth]{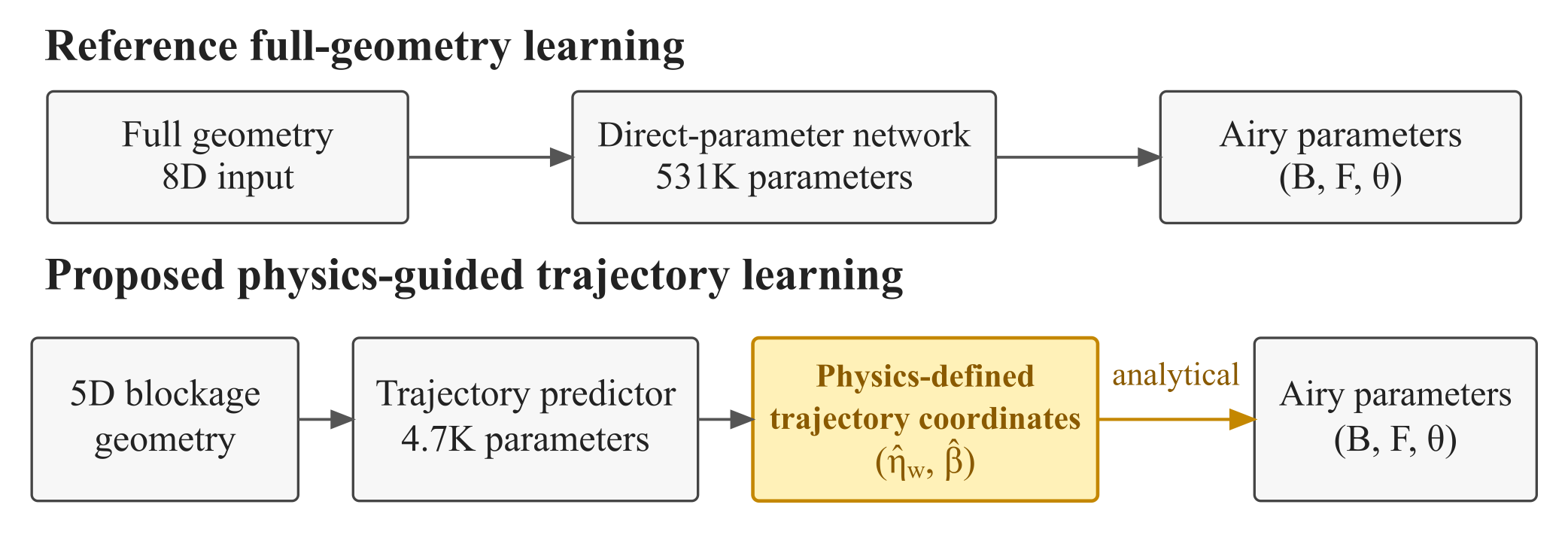}
\caption{Comparison of the reference full-geometry neural network and the proposed physics-guided neural predictor.}
\label{fig:architecture_flow}
\end{figure}

Figure~\ref{fig:architecture_flow} contrasts this structure with the existing full-geometry Airy learning architecture~\cite{chenhz}, which maps full-geometry features directly to $(B,F,\theta)$. The proposed architecture learns two trajectory coordinates from the stationary/KKT reference and leaves Airy generation to the analytical map. This separation permits a substantially smaller predictor. The full-geometry comparison in Section~\ref{sec:simulations} evaluates an adapted direct-parameter implementation of the reference architecture under the same propagation model and scenes.

\subsection{Online Beamforming Procedure}

From the perspective of an operational near-field link, the online procedure in Algorithm~\ref{alg:oneshot} runs as follows.

\begin{algorithm}[H]
\caption{Physics-Guided One-Shot Airy Beamforming}
\label{alg:oneshot}
\footnotesize
\begin{algorithmic}[1]
\REQUIRE Receiver geometry $(z_r,x_r)$, obstacle edge $(z_o,x_e)$, unobstructed-side sign $s$, trained predictor $f_{\bm\vartheta}$, normalization statistics, and chart $\mathcal Q$.
\STATE Compute $\rho$ using \eqref{eq:blockage_ratio_definition} and form $\mathbf g$ using \eqref{eq:predictor_input}.
\STATE Standardize $\mathbf g$ and predict $(\hat\eta_w,\hat\beta)$ using \eqref{eq:predictor_output}.
\STATE De-standardize and bound $(\hat\eta_w,\hat\beta)$ by the fixed numerical chart $\mathcal Q$.
\STATE Construct $(z_w,x_w)$ from $(\hat\eta_w,\hat\beta)$ using \eqref{eq:waypoint_coordinates}.
\STATE Obtain $(B,F,\theta)$ using the analytical generation map \eqref{eq:waypoint_to_params}.
\STATE Generate the discrete $N$-element ULA beamforming vector $\bw$ using \eqref{eq:discrete_airy_beamforming_vector}.
\STATE Transmit $\bw$.
\end{algorithmic}
\end{algorithm}

The procedure takes as input the sensed scene geometry: the receiver position $(z_r,x_r)$, the obstacle edge location $(z_o,x_e)$, and the unobstructed-side sign $s$. These five quantities come from sensing, environment perception, or a higher-layer map, so no complete post-blockage CSI or receiver-feedback sweeping is required.

Step~1 derives the aperture blockage ratio $\rho$ from \eqref{eq:blockage_ratio_definition}, which measures the blocked aperture fraction as $\rho=1-|\mathcal X_{\rm vis}|/|\mathcal X_{\rm tx}|$. Because $\rho$ follows from the known blockage geometry and aperture, it adds no extra sensing burden; it is then appended to the five geometry quantities to form the six-dimensional feature $\mathbf g=[z_r,x_r,z_o,x_e,s,\rho]^{\mathrm T}$ in \eqref{eq:predictor_input}.

Step~2 feeds the standardized $\mathbf g$ to the trained MLP and obtains the continuous trajectory coordinates $(\hat\eta_w,\hat\beta)=f_{\bm\vartheta}(\mathbf g)$ from \eqref{eq:predictor_output}. The predictor has learned the nonlinear blockage-geometry-to-trajectory map offline and performs no propagation integral or numerical optimization online.

Step~3 de-standardizes the two outputs and clips them to the fixed numerical chart $\mathcal Q$, a safeguard that keeps the prediction within the coordinate range used to numerically realize the compact trajectory region in \eqref{eq:stationary_trajectory_region}.

Step~4 reconstructs the physical waypoint $(z_w,x_w)$ from the bounded coordinates through \eqref{eq:waypoint_coordinates}, i.e., $z_w=z_o+\hat\beta(z_r-z_o)$ and $x_w=x_e+s\hat\eta_w r_F$. The coordinate $\hat\beta$ places the waypoint between the obstacle and receiver planes, while $\hat\eta_w$ specifies its transverse displacement from the obstacle edge.

Step~5 applies the closed-form generation operator \eqref{eq:waypoint_to_params}, $(B,F,\theta)=\mathcal G_{\rm A}(z_w,x_w;z_r,x_r,s)$, where the sign $s$ selects the bending branch, to obtain the Airy control triplet without any candidate search.

Step~6 evaluates the discrete beamforming vector from \eqref{eq:discrete_airy_beamforming_vector}, $\mathbf w(B,F,\theta)=[\tilde w_1,\ldots,\tilde w_N]^{\mathrm T}/\|\cdot\|_2$ with $\tilde w_n=u_0(x_n;B,F,\theta)$, realizing the selected trajectory on the $N$-element array.

Step~7 transmits the single beam $\mathbf w$. All propagation integrals, stationary/KKT reference calculations, and receiver-window power evaluations remain offline and do not enter this online chain.

\subsection{Complexity and Overhead}

For an $L_{\rm NN}$-layer MLP of widths $\{d_\ell\}_{\ell=0}^{L_{\rm NN}}$, the neural inference cost is $\mathcal O(\sum_{\ell=1}^{L_{\rm NN}}d_{\ell-1}d_\ell)$. The coordinate and Airy-control maps have constant cost, while generating the final $N$-element beamforming vector has $\mathcal O(N)$ cost. In contrast to candidate-based beam training, only one beam is transmitted. Because the physics-guided representation expresses trajectory selection using two coordinates associated with the compact trajectory region, the predictor can use a small network. The online cost is therefore dominated by a few matrix--vector products rather than iterative optimization or beam sweeping. This overhead statement excludes the separate cost of geometry sensing or map acquisition.

\section{Simulation Results}\label{sec:simulations}

The simulation results validate the compact physics-defined region and evaluate the proposed one-shot framework in terms of achievable rate, beam-training overhead, and parameter efficiency.

\subsection{Simulation Setup}

All simulations use the single-edge finite-aperture scalar Fresnel model in Section~\ref{sec:system} at $f_c=140$ GHz and $W=1$ GHz with a 256-element half-wavelength-spaced ULA. All beamforming vectors have equal transmit power, and the receiver-window average power in \eqref{eq:rx_power} is the common physical metric. We consider the partial-blockage regime $\rho\in[\rho_{\min},\rho_{\max})$, chosen to exclude the LoS-dominated and near-total-blockage limits. Complete scene ranges, data splits, numerical and training settings, and feasibility diagnostics are reported in Appendix~\ref{app:simulation_details}.

The numerical chart $\mathcal Q$, Fresnel threshold $\tau_{\rm F}$, and tolerance $\epsilon_{\rm dB}=0.5$ dB are fixed before independent testing. The tolerance defines both the competitive stationary set in \eqref{eq:competitive_stationary_set} and the local curvature width in \eqref{eq:curvature_width}. For $W=1$ GHz, it bounds the corresponding rate loss by $0.166$ Gbps at any SNR. During inference, $\mathcal Q$ only bounds the predictor outputs, whereas $\tau_{\rm F}$ and $\epsilon_{\rm dB}$ do not enter the neural forward pass or the rate formula.

For method $m$, the scene-wise mean rate over $N_{\rm te}$ test scenes is
\begin{equation}
	\bar R_m=\frac{1}{N_{\rm te}}\sum_{i=1}^{N_{\rm te}}
	W\log_2\!\left(1+\gamma_{\rm ref}
	\frac{P_{m,i}}{P_{{\rm free-focus},i}}\right),
\label{eq:scene_wise_rate}
\end{equation}
where $P_{m,i}$ is the blocked receiver-window power in scene $i$, $P_{{\rm free-focus},i}$ is the corresponding focused free-space power, and $\gamma_{\rm ref}=10^{30/10}$. This metric is normalized to the focused free-space benchmark rather than absolute capacity, and $\alpha_0$ cancels in the power ratio. Since achievable rate is strictly increasing in received power, $\gamma_{\rm ref}$ does not alter the power-maximizing trajectory.

\subsection{Validation of the Compact Physics-Defined Region}\label{sec:region_validation}

The ability of the compact physics-defined region to retain near-optimal trajectories while substantially reducing the candidate-space area is evaluated using 72 independent scenes stratified by blockage ratio, obstacle depth, and unobstructed side. The broad waypoint reference is obtained offline by numerical optimization over the full waypoint range $\mathcal Q$ and serves only as a validation benchmark. Table~\ref{tab:region_validation_by_blockage} reports the broad-reference coverage and retained candidate-space area across the three blockage intervals.

\begin{table}[H]
\centering
\caption{Validation of the compact physics-defined region across blockage intervals.}
\label{tab:region_validation_by_blockage}
\footnotesize
\setlength{\tabcolsep}{3pt}
\renewcommand{\arraystretch}{1.05}
\begin{tabular}{cccc}
\hline
$\rho$ interval & Scenes & \shortstack{Broad-reference\\coverage} & \shortstack{Mean retained\\area (\%)} \\
\hline
$[0.50,0.62)$ & 24 & $100\%$ & $2.12$ \\
$[0.62,0.74)$ & 24 & $100\%$ & $6.67$ \\
$[0.74,0.86)$ & 24 & $100\%$ & $9.15$ \\
Overall & 72 & $100\%$ & $5.98$ \\
\hline
\end{tabular}
\end{table}

At $\epsilon_{\rm dB}=0.5$ dB, Table~\ref{tab:region_validation_by_blockage} shows that the compact region achieves $100\%$ broad-reference coverage across all three blockage intervals while retaining $5.98\%$ of the candidate-space area on average. The retained area remains below $10\%$ in every interval, so the full waypoint range has approximately $16.7\times$ the area of the compact region. The offline trajectory reference $\mathbf q^\star$ in \eqref{eq:stationary_kkt_reference} supplies the predictor targets. It also matches the independently obtained broad waypoint reference to numerical precision, with a mean received-power gap of $3.22\times10^{-13}$ dB. Accordingly, $\mathbf q^\star$ is used as the broad waypoint reference for the full 360-scene evaluation. Detailed numerical checks are reported in Appendix~\ref{app:simulation_details}.

\subsection{One-Shot Performance}\label{sec:rate_overhead}

The rate--overhead benefit of replacing repeated Airy beam training with one-shot trajectory selection is evaluated on a 360-scene independent holdout. The proposed networks with 4,740 and 34,178 trainable parameters are denoted the 4.7K and 34K predictors, respectively.

\begin{figure}[!t]
	\centering
	\includegraphics[width=\columnwidth]{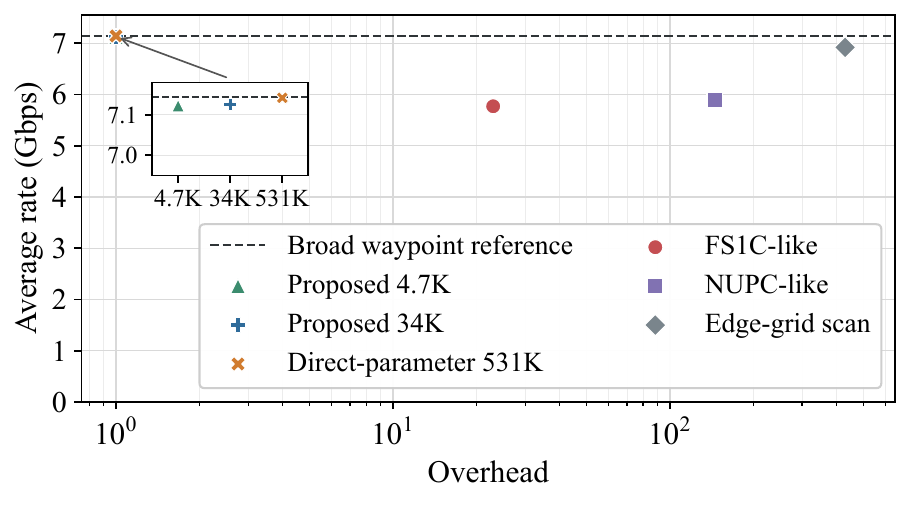}
	\caption{Mean scene-wise achievable rate versus transmitted-beam overhead over 360 independent test scenes.}
	\label{fig:rate_overhead}
\end{figure}

The beam-training baselines are adapted from~\cite{zhao2026efficienttraining}. The fast-scanning one-dimensional codebook (FS1C-like) baseline tests 21 lateral candidates plus two probes, for a total of 23 transmitted beams. The non-uniform polar codebook (NUPC-like) baseline tests a $13\times11$ waypoint grid plus two probes, for a total of 145 beams. The edge-grid scan transmits $33\times13$ candidates without separate probes, for a total of 429 beams. Each neural predictor instead selects and transmits one beam. Figure~\ref{fig:rate_overhead} compares the resulting rates and transmitted-beam overhead, with the broad waypoint reference included only as an offline benchmark.

Figure~\ref{fig:rate_overhead} shows that both proposed predictors approach the broad waypoint reference with one transmitted beam. The inset resolves their closely clustered results. The 4.7K predictor reaches $7.123$ Gbps and retains $99.70\%$ of the $7.144$-Gbps broad waypoint reference. Increasing the predictor size to 34K adds only $0.003$ Gbps. By comparison, the finite scans require 23--429 beams and reach $5.769$--$6.918$ Gbps. Together, the two evaluations distinguish region construction from one-shot prediction: Section~\ref{sec:region_validation} verifies reference coverage on independent validation scenes, while Fig.~\ref{fig:rate_overhead} measures the predictor rate gap on the 360-scene holdout.

\begin{figure}[!t]
	\centering
	\includegraphics[width=\columnwidth]{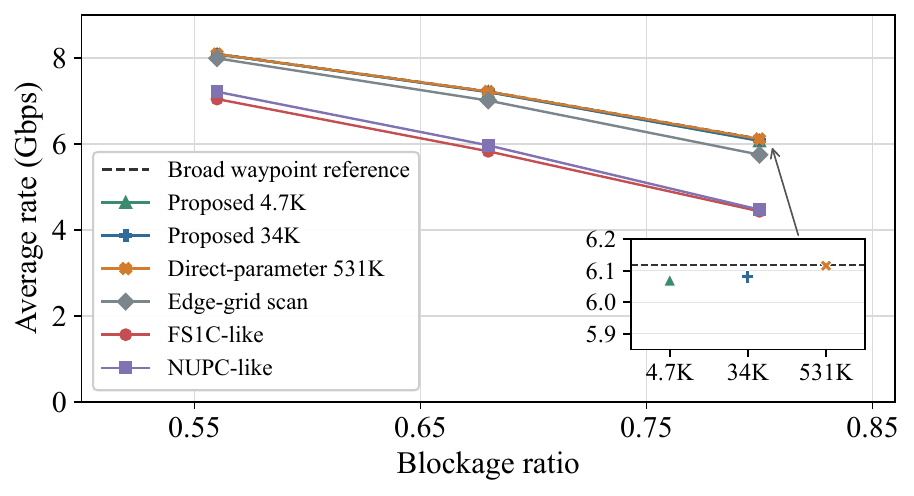}
	\caption{Mean scene-wise achievable rate versus blockage ratio over 360 independent test scenes.}
	\label{fig:rate_blockage}
\end{figure}

Figure~\ref{fig:rate_blockage} further examines whether the performance advantage persists as blockage becomes stronger. The inset in Fig.~\ref{fig:rate_blockage} resolves the one-shot results in the highest blockage interval. The same ordering holds across all evaluated blockage intervals. Even in the highest interval, the 4.7K predictor achieves $6.070$ Gbps versus $6.118$ Gbps for the broad waypoint reference, retaining $99.22\%$ of the reference rate. Meanwhile, the edge-grid gap increases with blockage. Complete numerical results are reported in Appendix~\ref{app:simulation_details}.

\subsection{Parameter-Efficiency Comparison}\label{sec:param_efficiency}

\begin{figure}[!t]
	\centering
	\includegraphics[width=\columnwidth]{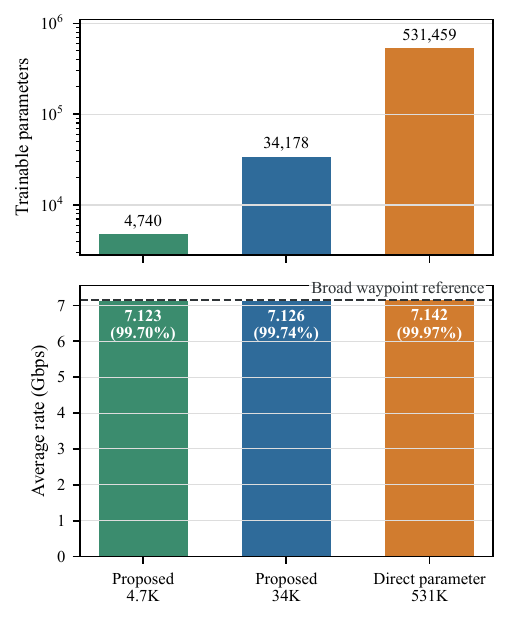}
	\caption{Trainable-parameter count (top) and mean scene-wise achievable rate (bottom) for the three one-shot methods.}
	\label{fig:neural_efficiency}
\end{figure}

Finally, the parameter efficiency of the compact physics-defined representation is evaluated against a data-driven full-geometry baseline adapted from~\cite{chenhz}. The resulting 531K-parameter network maps an eight-dimensional blockage geometry directly to the Airy control triplet $(B,F,\theta)$. In contrast, the proposed predictor learns the two trajectory coordinates associated with the compact region and obtains the Airy controls through the analytical generation map. All three neural methods use the same propagation model, data splits, and checkpoint-selection criterion. Each method transmits one beam. Because the data-driven baseline is not restricted to the compact trajectory region, the broad waypoint reference is a comparison benchmark rather than a strict upper bound. Figure~\ref{fig:neural_efficiency} compares their model sizes and rates.

Figure~\ref{fig:neural_efficiency} shows that the proposed 4.7K predictor nearly matches the 531K data-driven network in achievable rate. In the lower panel, the dashed line denotes the broad waypoint reference rate, and the bar labels report rate retention. The data-driven 531K network reaches $7.142$ Gbps, only $0.019$ Gbps above the proposed 4.7K predictor. Thus, the proposed physics-guided neural predictor retains more than $99.7\%$ of the data-driven network rate with approximately $112\times$ fewer trainable parameters. The compact physics-defined representation therefore reduces the required neural model size without a material rate loss. The full training recipe is given in Appendix~\ref{app:simulation_details}.

\section{Conclusions}\label{sec:conclusion}

This paper developed a physics-guided neural framework for one-shot Airy beamforming under near-field blockage. We first revealed the trajectory--edge coupling mechanism and derived the corresponding optimality condition for trajectory selection. Based on the optimality condition, a continuous trajectory structure was established and exploited to construct a compact physics-defined region. A physics-guided neural predictor was then designed to select a near-optimal trajectory in one shot. Simulations show that the compact physics-defined region effectively captures near-optimal trajectories. The proposed framework outperforms conventional beam-training methods and reduces the beam-training overhead to one transmission. Moreover, the lightweight physics-guided framework retains $99.7\%$ of the rate obtained through numerical optimization, and achieves performance comparable to the data-driven method with approximately $112\times$ fewer neural-network parameters.
Notably, the absence of a closed-form rule for selecting a beam trajectory under blockage is not unique to Airy beams and may also arise in other blockage-mitigation schemes. In such schemes, trajectory--edge coupling mechanism revealed in this paper can be used to identify a compact, physically structured candidate region, thereby reducing the candidate range for numerical optimization, beam training, or learning-based selection.
Future work can extend the proposed physics-guided Airy beamforming framework to multiple edges and uniform planar arrays, as well as to scenarios under sensing uncertainty and hardware constraints.

\FloatBarrier
\appendices
\section{Proof of Proposition 1}\label{app:proof_stationary_branch}

Let $G(\eta_w,\beta)=P_{\eta_w}(\eta_w,\beta)$. At $(\eta_j,\beta_0)$, $G=0$ and $G_{\eta_w}=P_{\eta_w\eta_w}<0$. The implicit-function theorem therefore gives a unique local branch $\eta_j(\beta)$, and differentiating $G[\eta_j(\beta),\beta]=0$ gives \eqref{eq:stationary_branch_slope}. Since $P_{\rm rx}>0$ and $P_{\eta_w}=0$ on the branch,
\begin{equation}
\frac{\partial^2\ln P_{\rm rx}}{\partial\eta_w^2}
=\frac{P_{\eta_w\eta_w}}{P_{\rm rx}}<0,
\label{eq:log_power_curvature}
\end{equation}
which establishes $\kappa_j>0$. Taylor's theorem then gives
\begin{equation}
\ln P_{\rm rx}(\eta_j+\delta_\eta,\beta)
=\ln P_{\rm rx}(\eta_j,\beta)
-\frac{\kappa_j}{2}\delta_\eta^2
+\mathcal O(|\delta_\eta|^3).
\label{eq:log_power_taylor}
\end{equation}
Multiplying the log-power difference by $10/\ln 10$ yields \eqref{eq:local_power_loss}, and solving its quadratic term for an $\epsilon_{\rm dB}$ loss gives \eqref{eq:curvature_width}. \hfill$\square$

\section{Fresnel Feasibility Test}\label{app:fresnel_feasibility}

For propagation distance $L$ and normalized transverse slope $v$, the phase omitted by the quadratic range expansion is
\begin{equation}
\varepsilon_{\rm F}(v,L)
=kL\left(1+\frac{v^2}{2}-\sqrt{1+v^2}\right).
\label{eq:fresnel_phase_remainder}
\end{equation}
The maximum remainder is evaluated over the aperture-to-generation, generation-to-receiver, transmitter-to-edge, and edge-to-receiver segments. Their normalized slopes are
\begin{align}
v_{\rm tx}&=\max_{|x_0|\leq D/2}
\left|\frac{1}{k}\frac{\partial\phi_{\rm A}(x_0;B,F,\theta)}
{\partial x_0}\right|,\nonumber\\
v_{\rm A}&=\max_{0\leq\ell\leq L_{\rm A}}
\left|\frac{\mathrm d x_{\rm A}(\ell)}{\mathrm d\ell}\right|,\nonumber\\
v_{o,\rm tx}&=\frac{|x_e|+D/2}{z_o},\nonumber\\
v_{o,\rm rx}&=\frac{|x_r-x_e|}{z_r-z_o},
\label{eq:segment_slopes}
\end{align}
where $x_{\rm A}(\ell)$ is the closed-form Airy trajectory in the generation-plane coordinate. The maximum phase remainder is
\begin{equation}
\begin{aligned}
\varepsilon_{\max}(\eta_w,\beta;\mathcal I)=\max\{&
\varepsilon_{\rm F}(v_{\rm tx},F),
\varepsilon_{\rm F}(v_{\rm A},L_{\rm A}),\\
&\varepsilon_{\rm F}(v_{o,\rm tx},z_o),
\varepsilon_{\rm F}(v_{o,\rm rx},z_r-z_o)\}.
\end{aligned}
\label{eq:maximum_fresnel_remainder}
\end{equation}
The simulations use $\tau_{\rm F}=0.5$ rad as a conservative sub-radian upper bound on each omitted phase term in \eqref{eq:maximum_fresnel_remainder}.

\section{Simulation and Training Details}\label{app:simulation_details}

Table~\ref{tab:simulation_setup} summarizes the common simulation and training configuration. Table~\ref{tab:region_validation} collects the numerical checks underlying the compact physics-defined region validation in Section~\ref{sec:simulations}, and Table~\ref{tab:rate_overhead} reports the complete rate and overhead results used in Figs.~\ref{fig:rate_overhead} and~\ref{fig:rate_blockage}.

\begin{table}[!b]
	\centering
	\caption{Simulation and training configuration.}
	\label{tab:simulation_setup}
	\footnotesize
	\renewcommand{\arraystretch}{1.08}
	\begin{tabular}{p{0.27\columnwidth}p{0.60\columnwidth}}
		\hline
		Group & Setting \\
		\hline
		Communication and array & $f_c=140$ GHz, $W=1$ GHz, $N=256$, $d=\lambda/2$, $D=0.273$ m, $\omega_0=D/2$, and $W_r=0.01$ m \\
		Scene geometry & $z_r\in[2.5,4.0]$ m, $x_r\in[0.04,0.12]$ m, $z_o/z_r\in[0.42,0.66)$, $\rho\in[\rho_{\min},\rho_{\max})$ with $\rho_{\min}=0.50$ and $\rho_{\max}=0.86$, and $s\in\{-1,+1\}$ \\
		Data split & 704 training scenes including 200 additional scenes from a sparsely sampled high-blockage region, 72 validation scenes, and 360 independently generated test scenes \\
		Offline reference & Exact stationary/KKT roots in $\eta_w\in[-4,4]$ and $\beta\in[0,0.95]$, $\tau_{\rm F}=0.5$ rad, $\epsilon_{\rm dB}=0.5$ dB for the competitive stationary set and curvature width, and independent broad-search validation \\
		Predictor training & SiLU MLPs with three hidden layers of width 46 or 128, 4,740/34,178 parameters, AdamW with constant learning rate $10^{-3}$ and weight decay $10^{-5}$, batch size 64, normalized SmoothL1, 3000 epochs, validation-power checkpoint selection, and seeds 732--734 \\
		\hline
	\end{tabular}
\end{table}

The 200 additional high-blockage scenes are used only for training. The 360-scene independent holdout was generated after the augmentation rule and all proposed-predictor checkpoints had been frozen, and it is not used for checkpoint selection for any method. Extrapolation beyond the scene ranges in Table~\ref{tab:simulation_setup} is not evaluated.

For any SNR, a received-power loss of $\epsilon_{\rm dB}$ dB produces a rate loss no greater than $W\epsilon_{\rm dB}\log_2(10)/10$, which equals $0.166$ Gbps for $W=1$ GHz.

The fixed numerical chart $\mathcal Q$ bounds the predictor outputs, whereas $\mathcal F_{\rm A}(\mathcal I)$ characterizes consistency with the generation-map regularity and Fresnel assumptions. Across the independent holdout and three random seeds, every predicted coordinate pair produces a well-defined Airy control triplet. The Fresnel-remainder criterion $\tau_{\rm F}=0.5$ rad is met by $99.35\%$ and $99.17\%$ of the predictions from the 4.7K and 34K predictors, respectively. Their maximum phase remainders are $0.650$ and $0.712$ rad.

\begin{table}[!t]
\centering
\caption{Numerical validation of the compact physics-defined region.}
\label{tab:region_validation}
\footnotesize
\renewcommand{\arraystretch}{1.05}
\setlength{\tabcolsep}{3pt}
\begin{tabular}{lp{0.54\columnwidth}}
\hline
Validation & Result \\
\hline
		Composed Fresnel integral & Maximum absolute log-power error $8.88\times10^{-16}$ \\
		Exact waypoint gradient & p95 scaled finite-difference error $1.40\times10^{-5}$ \\
		Physics-defined region & 72 stratified scenes: $100\%$ broad-reference coverage and $5.98\%$ mean chart-area fraction \\
		Independent broad validation & Stationary/KKT-to-broad-reference mean gap $3.22\times10^{-13}$ dB over the same 72 scenes \\
\hline
\end{tabular}
\end{table}

For the full-geometry comparison, the eight-dimensional input is
$\mathbf g_{\rm fg}=[z_o,x_o,L_z,w_o,z_r,x_r,\rho,\Delta_x]^{\mathrm T}$.
An equivalent thin obstacle is constructed with $L_z=0.02$ m,
$w_o=(1-z_o/z_r)\rho D$, $x_o=x_e-sw_o/2$, and
$\Delta_x=x_r-(x_o/z_o)z_r$. This construction preserves the aperture blockage ratio and is used only to express the single-edge geometry in the input form of the reference architecture. The $3\times512$ ReLU MLP predicts $(B,\Delta F,\Delta\theta)$, where $\Delta F$ and $\Delta\theta$ are offsets from the focused range $\sqrt{z_r^2+x_r^2}$ and angle $\operatorname{atan2}(x_r,z_r)$, respectively. It has 531,459 trainable parameters and follows the reference recipe~\cite{chenhz} with SGD, a learning rate of $0.008$, momentum of $0.8$, dropout of $0.15$, batch size 64, and normalized parameter L1 loss for 1000 epochs. Three models with seeds 742--744 are evaluated. Each architecture retains its established training recipe, and checkpoints for all architectures are selected according to validation receiver-window power.

For method $m$, the reported gap to the broad waypoint reference is the mean scene-wise power gap
$\bar G_m=N_{\rm te}^{-1}\sum_{i=1}^{N_{\rm te}}10\log_{10}(P_{{\rm ref},i}/P_{m,i})$.

\begin{table}[!t]
	\centering
	\caption{Scene-wise rate and online overhead on 360 independent test scenes. Neural results are mean $\pm$ standard deviation over three seeds.}
	\label{tab:rate_overhead}
	\footnotesize
	\renewcommand{\arraystretch}{1.05}
	\setlength{\tabcolsep}{4pt}
	\begin{tabular}{lccc}
		\hline
		Method & Overhead & Rate (Gbps) & \shortstack{Gap to broad\\ref. (dB)} \\
		\hline
		FS1C-like & 23 & 5.769 & 4.213 \\
		NUPC-like & 145 & 5.885 & 3.860 \\
		Edge-grid scan ($33\times13$) & 429 & 6.918 & 0.692 \\
		Proposed 4.7K & 1 & $7.123\pm0.007$ & $0.066\pm0.021$ \\
		Proposed 34K & 1 & $7.126\pm0.007$ & $0.057\pm0.022$ \\
		Direct-parameter 531K & 1 & $7.142\pm0.0005$ & $0.007\pm0.001$ \\
		Broad waypoint reference & Offline & 7.144 & 0 \\
		\hline
	\end{tabular}
\end{table}

\FloatBarrier

\footnotesize
\bibliographystyle{IEEEtran}
\bibliography{ref}

@book{2021THz,
	title={{THz} Communications: Paving the Way Towards Wireless {Tbps}},
	editor={Thomas K{\"u}rner and Daniel M. Mittleman and Tadao Nagatsuma},
	series={Springer Series in Optical Sciences},
	volume={234},
	address={Cham, Switzerland},
	publisher={Springer},
	year={2022},
}

@book{optics,
	title={Introduction to {Fourier} Optics},
	author={ Goodman, Joseph W },
	address={San Francisco, CA, USA},
	publisher={McGraw-Hill},
	year={1968},
}

@misc{3gpp38901,
	author={{3rd Generation Partnership Project (3GPP)}},
	title={Study on Channel Model for Frequencies from 0.5 to 100 {GHz}},
	howpublished={3GPP TR 38.901, V19.4.0},
	year={2026},
	month={Jun.},
}

@techreport{iturp526,
	author={{ITU-R}},
	title={Propagation by Diffraction},
	institution={International Telecommunication Union},
	address={Geneva, Switzerland},
	type={Recommendation},
	number={ITU-R P.526-16},
	year={2025},
	month={Nov.},
}

@article{airy1,
	author = {Berry, M. V. and Balazs, N. L.},
	title = {Nonspreading wave packets},
	journal = {Am. J. Phys.},
	volume = {47},
	number = {3},
	pages = {264-267},
	year = {1979},
	month = {Mar.},
	issn = {0002-9505},
	doi = {10.1119/1.11855},
}

@article{airyfinite,
	title={Accelerating finite energy {Airy} beams},
	author={Siviloglou, Georgios A. and Christodoulides, Demetrios N.},
	journal={Opt. Lett.},
	volume={32},
	number={8},
	pages={979--981},
	year={2007},
	month={Apr.},
	doi={10.1364/OL.32.000979},
	publisher={Optica Publishing Group}
}

@article{airy5,
	title={Observation of accelerating {Airy} beams},
	author={Siviloglou, Georgios A and Broky, Jone and Dogariu, Aristide and Christodoulides, DN},
	journal={Phys. Rev. Lett.},
	volume={99},
	number={21},
	note={{Art. no. 213901}},
	year={2007},
	month={Nov.},
	publisher={APS}
}

@article{overview,
	title={{Airy} beams and accelerating waves: an overview of recent advances},
	author={Efremidis, Nikolaos K and Chen, Zhigang and Segev, Mordechai and Christodoulides, Demetrios N},
	journal={Optica},
	volume={6},
	number={5},
	pages={686--701},
	year={2019},
	month={May},
	publisher={Optical Society of America}
}

@article{selfhealing1,
	title={Self-healing properties of optical {Airy} beams},
	author={Broky, John and Siviloglou, Georgios A and Dogariu, Aristide and Christodoulides, Demetrios N},
	journal={Opt. Express},
	volume={16},
	number={17},
	pages={12880--12891},
	month={Aug.},
	year={2008},
	publisher={Optical Society of America}
}

@article{selfhealing2,
	title={Analytical study of the self-healing property of {Airy} beams},
	author={Chu, Xiuxiang and Zhou, Guoquan and Chen, Ruipin},
	journal={Phys. Rev. A},
	volume={85},
	number={1},
	note={{Art. no. 013815}},
	month={Jan.},
	year={2012},
	publisher={APS}
}

@article{curving,
	title={Curving {THz} wireless data links around obstacles},
	author={Guerboukha, Hichem and Zhao, Bin and Fang, Zhaoji and Knightly, Edward and Mittleman, Daniel M},
	journal={Commun. Eng.},
	volume={3},
	month={Mar.},
	note={{Art. no. 58}},
	year={2024},
	publisher={Nature Publishing Group UK London}
}

@article{lee,
	title={Experimental Demonstration of Wireless Transmission Using {Airy} Beams in {Sub-THz} Band},
	author={Lee, Doohwan and Yagi, Yasunori and Suzuoki, Kosuke and Kudo, Riichi},
	journal={IEEE Open J. Commun. Soc.},
	volume={6},
	pages={1091--1102},
	year={2025},
	month={Jan.},
	doi={10.1109/OJCOMS.2025.3534788},
	publisher={IEEE}
}

@article{chenhz,
	title={A physics-informed {Airy} beam learning framework for blockage avoidance in sub-terahertz wireless networks},
	author={Chen, Haoze and Kludze, Atsutse and Ghasempour, Yasaman},
	journal={Nat. Commun.},
	volume={16},
	month={Aug.},
	note={{Art. no. 7387}},
	year={2025},
}

@ARTICLE{hanchong,
	author={Zhao, Wenqi and Abadal, Sergi and Song, Guochao and Jiang, Jiamo and Han, Chong},
	journal={IEEE Trans. Wireless Commun.}, 
	title={{Terahertz} Wireless Data Center: {Gaussian} Beam or {Airy} Beam?}, 
	year={2026},
	volume={25},
	number={},
	pages={7922--7938},
	doi={10.1109/TWC.2025.3634516}}

@misc{hanchong2,
	title={{Airy} Beam Engineering in Near-field Communications: A Tractable Closed-Form Analysis in the {Terahertz} Band}, 
	author={Wenqi Zhao and Chong Han and Emil Bj{\"o}rnson},
	howpublished={arXiv:2603.13866},
	year={2026},
}

@misc{zhao2026efficienttraining,
	title={Efficient {Airy} Beam Training for Quasi-{LoS} {Terahertz} Near-Field Communications},
	author={Zhao, Wenqi and Han, Chong},
	howpublished={arXiv:2605.09895},
	year={2026},
}

@misc{songlingyang,
	title={Breaking Near-Field Communication Barriers: Focused, Curved, or {Airy} Beamforming?}, 
	author={Shupei Zhang and Boya Di and Lingyang Song},
	howpublished={arXiv:2604.01704},
	year={2026},
}

@misc{darsena2025airy,
	title={{Airy} Beams for Near-Field Communications: Fundamentals, Potentials, and Limitations},
	author={Darsena, Donatella and Verde, Francesco and Di Renzo, Marco and Galdi, Vincenzo},
	howpublished={arXiv:2508.13714},
	year={2025},
}

@MISC{galeote2026blockage,
	author={Galeote-Cazorla, Juan E. and Ram{\'i}rez-Arroyo, Alejandro and Molina-Garcia-Pardo, Jose-Maria and Martinez-Ingles, Maria-Teresa and Valenzuela-Vald{\'e}s, Juan F.},
	title={Experimental Assessment of Human Blockage at {sub-THz} and {mmWave} Frequency Bands},
	howpublished={\emph{IEEE Trans. Veh. Technol.}, early access},
	month={Apr. 8,},
	year={2026},
	note={doi: 10.1109/TVT.2026.3682016},
}

@ARTICLE{XLMIMO2,
  author={Cui, Mingyao and Wu, Zidong and Lu, Yu and Wei, Xiuhong and Dai, Linglong},
  journal={IEEE Commun. Mag.}, 
  title={Near-Field {MIMO} Communications for {6G}: Fundamentals, Challenges, Potentials, and Future Directions}, 
  year={2023},
  volume={61},
  number={1},
  pages={40-46},
  month= {Jan.},
}

@misc{wang2026multiairy,
	title={Blockage-Robust Beamforming for Near-Field Communications: From Single-{Airy} to Multi-{Airy}},
	author={Wang, Yi and Dai, Linglong},
	howpublished={arXiv:2607.07278},
	year={2026},
}

@book{rudin1976principles,
	author={Rudin, Walter},
	title={Principles of Mathematical Analysis},
	edition={3rd},
	publisher={McGraw-Hill},
	address={New York, NY, USA},
	year={1976},
}

\end{document}